\documentclass[iop]{emulateapj}

\usepackage{graphicx}
\usepackage{amsmath}
\usepackage{appendix}
\usepackage{amssymb}
\usepackage{aas_macros}
\usepackage{natbib}
\usepackage{latexsym}
\usepackage{longtable}
\usepackage{threeparttable}
\usepackage{subfigure}
\usepackage[figoff]{figcaps}
\usepackage{ccaption}
\captiondelim{ }

\begin{document}

% %%%%%%%%%%%%%%%%%%%%%%%%%%%%%%%%%%%
\pagestyle{plain}
\title{Predicting Planets in {\em Kepler} Multi-Planet Systems}

\author{Julia Fang\altaffilmark{1} and Jean-Luc Margot\altaffilmark{1,2}}

\altaffiltext{1}{Department of Physics and Astronomy, University of California, Los Angeles, CA 90095, USA}
\altaffiltext{2}{Department of Earth and Space Sciences, University of California, Los Angeles, CA 90095, USA}

\begin{abstract}

We investigate whether any multi-planet systems among {\em Kepler}
candidates (2011 February release) can harbor additional
terrestrial-mass planets or smaller bodies.  We apply the ``packed
planetary systems'' hypothesis that suggests all planetary systems are
filled to capacity, and use a Hill stability criterion to identify
eight 2-planet systems with significant gaps between the innermost and
outermost planets. For each of these systems, we perform long-term
numerical integrations of 10$^7$ years to investigate the stability of
4000$-$8000 test particles injected into the gaps. We map out
stability regions in orbital parameter space, and therefore quantify
the ranges of semi-major axes and eccentricities of stable particles.
Strong mean-motion resonances can add additional regions of stability
in otherwise unstable parameter space.  We derive simple expressions
for the extent of the stability regions, which is related to
quantities such as the dynamical spacing $\Delta$, the separation
between two planets in units of their mutual Hill radii.  Our results
suggest that planets with separation $\Delta < 10$ are unlikely to
host extensive stability regions, and that about 95 out of a total of
115 two-planet systems in the {\em Kepler} sample may have sizeable
stability regions.  We predict that {\em Kepler} candidate systems
including KOI 433, KOI 72/Kepler-10, KOI 555, KOI 1596, KOI 904, KOI
223, KOI 1590, and KOI 139 can harbor additional planets or low-mass
bodies between the inner and outer detected planets. These predicted
planets may be detected by future observations.

\end{abstract}
\maketitle

% %%%%%%%%%%%%%%%%%%%%%%%%%%%%%%%%%%%
\section{Introduction}

Early studies of extrasolar planetary systems showed residual velocity
trends in Keplerian orbit fits to radial velocity data
\citep[e.g.,][]{marc98,butl98,marc99,vogt00,fisc01}, suggesting that
these systems may host additional, undetected
planets. \citet{fisc01} noted that about half of the stars in their
sample of 12 systems showed residual trends greater than the expected
scatter due to measurement uncertainties and stellar noise. Most of
these systems were later confirmed to harbor additional planet(s).

In more recent years, the study and prediction of undiscovered planets
have been aided by long-term N-body simulations. These numerical
investigations searched for stability zones in multi-planet systems by
integrating hundreds to thousands of test bodies, which were injected
into empty regions between known planets
\citep[e.g.,][]{meno03,barn04,raym05,ji05,rive07,raym08}. 
For example, a putative Saturn-mass planet in HD~74156 was first
predicted by \citet{raym05} through numerical simulations that showed
a stable region between planets b and c.  The planet was later
discovered by \citet{bean08}, although there have been questions about
the validity of the detection~\citep{witt09,mesc11}.
This prediction was motivated by the ``packed planetary systems''
(PPS) hypothesis.

The PPS hypothesis is the idea that planetary systems are formed
``dynamically full'' and filled to capacity, and any additional
planets will cause the systems to be unstable \citep[e.g.,][]{barn04,
raym05, raym06, barn07}. Consequently, planetary systems with stable
stability zones between the innermost and outermost planets are
suggestive of additional, yet-undetected planets. Reasons for the
non-detections of hypothetical planets include lack of sufficient
data, such as non-transiting planets that require more data to detect
them via transit timing variations, or planetary masses/radii that are
below detection limits. The orbital properties of predicted planets
can be identified through long-term numerical simulations. Support for
the PPS hypothesis comes from early observations of packed
multi-planet systems that led to this hypothesis
\citep[e.g.,][]{butl99,marc01b,marc01a,fisc02,mayo04}, apparent
consistency between the planet-planet scattering model and packed
systems \citep{raym09}, the remarkably dense and packed Kepler-11
system \citep{liss11a}, theoretical work
\citep[e.g.,][]{cham96,smit09}, and other investigations
\citep[e.g.,][]{rive00,gozd06}.

In the present study, we apply the PPS hypothesis to multi-planet
candidate systems discovered by the {\em Kepler} team during the
mission's first four and a half months of data \citep{boru11}. The
{\em Kepler} mission is a transit survey designed to search for
Earth-sized planets \citep{boru10,koch10,jenk10,cald10}, and is
sensitive to terrestrial-class and larger planets located at a large
range of separations from their host star. {\em Kepler} can detect
multiple transiting systems for densely-packed planets with nearly
coplanar configurations or with serendipitous geometric alignment, and
the dynamics and statistics of {\em Kepler} multi-planet systems are
providing a wealth of information about planetary systems
\citep[e.g.,][]{stef10,lath11,liss11a,liss11b}.

Given that planetary systems have been discovered with densely packed
planets, we seek to test the PPS hypothesis and predict additional
planets in {\em Kepler} candidate multi-planet systems. In Section
\ref{datamethods}, we discuss {\em Kepler}'s sample of multi-planet
systems as well as our methodology for evaluating each planetary
system's level of dynamical packing. We also explain our methods for
running numerical simulations and our choice of initial conditions. In
Section \ref{results}, we present the results from long-term numerical
integrations and illustrate them using stability maps. Section
\ref{mountain} discusses the dynamical interpretation of our work, in
particular the relationship between planetary spacing and the extent
of an inter-planet stability region.  We then summarize the
restrictions and scope of our study (Section \ref{scope}) and state
our conclusions (Section \ref{conclusion}).

% %%%%%%%%%%%%%%%%%%%%%%%%%%%%%%%%%%%
\section{Data and Methods} \label{datamethods}

Based on publicly-available {\em Kepler} data covering the first four
and a half months of observations, about one-third of $\sim$1200 
transiting planet candidates are hosted in multi-planet systems
\citep{boru11,liss11b}.  These multi-planet systems include 115
systems with 2 transiting planets, 45 systems with 3 transiting
planets, 8 systems with 4 transiting planets, 1 system with 5
transiting planets, and 1 system with 6 transiting planets. Most {\em
  Kepler} candidate planets have not been validated and are therefore
``Kepler Objects of Interest'' (KOI) and assigned a number. Candidates
in multi-planet systems have a smaller probability than
single-planet candidates of being an astrophysical false positive 
\citep{rago10,lath11,liss11a,liss12}. Moreover, all of these candidate multi-planet
systems are stable over long-term integrations \citep{liss11b}. 
For the remainder of this paper, we refer to all candidate planets 
and systems by dropping the adjective ``candidate.'' 
All of these {\em Kepler} multi-planet systems presented by
\citet{boru11} are examined using the analytical method described
below in Section \ref{analy}.

To discern the extent of packing in {\em Kepler} multi-planet systems,
we define two types of stability as outlined by
\citet{glad93}. Fulfillment or over-fulfillment of these stability
criteria, meaning that the considered planetary system is not on the
verge of instability, can imply the presence of additional planets
according to the PPS hypothesis. First, Hill stability requires that a
system's ordering of planets (in terms of distance from the star)
remains constant. This means that close approaches are forbidden and
planet-crossing is not allowed for all time, but the outer planet may
be unbound and still be Hill stable. The second type of stability is
Lagrange stability, which is a stricter definition than Hill
stability. Lagrange stability requires not only the conservation of
the ordering of planets, but also that they remain bound to the star
for all time. Hill stability can be mathematically examined for
two-planet, non-resonant systems, and Lagrange stability is typically
examined through numerical simulations. Hill stability can be a
reasonable predictor of Lagrange stability \citep{barn06}. In the next
two subsections, we examine Hill stability through analytical methods
(Section \ref{analy}) and Lagrange stability through N-body
integrations (Section \ref{numer}) for {\em Kepler} multi-planet
systems.

% %%%%%%%%%%%%%%%%%%%%%%%%%%%%%%%%%%%
\def\arraystretch{1.4}
\begin{deluxetable*}{lrrrrrrrrrrr}
\tablecolumns{12}
\tablecaption{Identified {\em Kepler} Systems with $\beta/\beta_{\rm crit} >$ 1.5 \label{betas}}
\startdata
\hline \hline
KOI	& $M_*$ ($M_{\odot}$) & $M_1$ ($M_{\earth}$) & $R_1$ ($R_{\earth}$) & $a_1$ (AU) & $P_1$ (days) & $M_2$ ($M_{\earth}$) & $R_2$ ($R_{\earth}$) & $a_2$ (AU) & $P_2$ (days) & $\beta/\beta_{\rm crit}$ & $\Delta$ \\
\hline
433        & 1.01 & 37.38 & 5.80 & 0.050 & 4.030 & 209.81 & 13.40 &  0.935 & 328.240 & 2.861 & 28.7 \\  
72$\dagger$& 1.03 & 1.72  & 1.30 & 0.018 & 0.837 & 5.56	  & 2.30  &  0.252 & 45.295  & 2.781 & 90.3 \\  
555        & 0.95 & 2.31  & 1.50 & 0.046 & 3.702 & 5.56	  & 2.30  &  0.376 & 86.496  & 2.031 & 77.3 \\  
1596       & 0.87 & 5.56  & 2.30 & 0.061 & 5.924 & 13.21  & 3.50  &  0.416 & 105.355 & 1.817 & 53.4 \\  
904        & 0.69 & 4.61  & 2.10 & 0.029 & 2.211 & 9.61	  & 3.00  &  0.159 & 27.939  & 1.624 & 50.4 \\  
223        & 0.92 & 7.74  & 2.70 & 0.041 & 3.177 & 6.07	  & 2.40  &  0.226 & 41.008  & 1.621 & 56.2 \\  
1590       & 0.88 & 3.75  & 1.90 & 0.033 & 2.356 & 8.34	  & 2.80  &  0.163 & 25.780  & 1.527 & 55.4 \\  
139        & 1.07 & 1.46  & 1.20 & 0.045 & 3.342 & 36.07  & 5.70  &  0.741 & 224.794 & 1.508 & 54.1 
\enddata

\tablenotetext{}{Eight {\em Kepler} multi-planet systems are
  identified with $\beta/\beta_{\rm crit}$ values greater than 1.5. We
  list their KOI (``Kepler Objects of Interest'') identifier, stellar
  mass $M_*$, planetary masses $M_1$ and $M_2$, planetary radii $R_1$
  and $R_2$, semi-major axes $a_1$ and $a_2$, orbital periods $P_1$
  and $P_2$, $\beta/\beta_{\rm crit}$ value (listed in descending
  order), and dynamical spacing criterion $\Delta$. The inner planet 
  has the subscript $1$ and the outer planet
  has the subscript $2$. Stellar mass and planetary parameters (size,
  semi-major axis, and period) are taken from \citet{boru11}, and we
  derived the planetary masses using a power law: $M_i =
  (R_i/R_{\earth})^{2.06} M_{\earth}$ where the subscript $i$
  represents planet 1 or 2 \citep{liss11b}. 
\\ $\dagger$ KOI 72 is a confirmed system and is also known as Kepler-10 \citep{bata11}.}

\end{deluxetable*}
% %%%%%%%%%%%%%%%%%%%%%%%%%%%%%%%%%%%

\subsection{Analytical Method} \label{analy}

We calculate the Hill stability criterion of each adjacent planet pair
in {\em Kepler} multi-planet systems. We follow the notation given in
\citet{barn06} by referring to the relevant quantities as $\beta$ and
$\beta_{\rm crit}$:
\begin{align} \label{eqn_beta}
	\beta = \dfrac{-2(M_* + M_1 + M_2)}{G^2 (M_1 M_2 + M_* M_1 + M_* M_2)^3} L^2 E
\end{align}
\begin{align} \label{eqn_betacrit}
	\beta_{\rm crit} = & 1 + 3^{4/3} \dfrac{M_1 M_2}{M_*^{2/3}(M_1 + M_2)^{4/3}} \\ 
	\nonumber & - \dfrac{M_1 M_2 (11M_1 + 7 M_2)}{3 M_* (M_1 + M_2)^2} + ...
\end{align}
where $M_*$ is the mass of the star, $M_1$ and $M_2$ are the masses of
the planets where $M_1>M_2$, $L$ and $E$ are the total orbital angular
momentum and energy of the system, and $G$ is the gravitational
constant \citep{marc82,glad93,vera04}. Two-planet, non-resonant
systems with $\beta/\beta_{\rm crit} \geq 1$ are considered Hill
stable. For systems with additional planets and/or in resonance,
stability needs to be investigated numerically. A system that does not
fulfill the Hill stability criterion has unknown Hill stability; it
may or may not be Hill stable. In Equations (\ref{eqn_beta}) and
(\ref{eqn_betacrit}), $\beta_{\rm crit}$ is only a function of masses
and $\beta$ is a function of masses as well as semi-major axes and
eccentricities; evidently, for a given set of masses, there are
stability boundaries in orbital parameter space.

We calculate $\beta/\beta_{\rm crit}$ for each adjacent planet pair in
{\em Kepler} multi-planet systems released as of February 2011 by 
\citet{boru11}. 
We find that almost all of the adjacent planet pairs have $\beta/\beta_{\rm crit}$
values greater than 1.
\citet{raym09} discussed that planet pairs with values of
$\beta/\beta_{\rm crit}$ greater than $\sim$1.5$-$2 are probably
capable of harboring additional planet(s) with a semi-major axis in
between those of the existing planets. We find eight {\em Kepler}
systems, all of which are two-planet systems, with $\beta/\beta_{\rm
  crit}$ values of 1.5 or greater (Table \ref{betas}). None of the
3-planet, 4-planet, 5-planet, and 6-planet systems have any adjacent
planet pairs with $\beta/\beta_{\rm crit}$ values of at least 1.5. In
the next section, we place test particles in each of these eight
systems to determine their zones of stability.

Another useful criterion for evaluating stability is the dynamical
spacing $\Delta$ between two planets, i.e.,\ the difference between
their semi-major axes expressed in units
of their mutual Hill radius,
\begin{align} \label{deltaeqn1}
	\Delta = \frac{a_2-a_1}{R_{H1,2}},
\end{align}
where $R_{H1,2}$ is the mutual Hill radius defined as
\begin{align} \label{deltaeqn2}
	R_{H1,2} = \left( \frac{M_1+M_2}{3M_*} \right)^{1/3}\frac{a_1+a_2}{2},
\end{align}
\citep[e.g.,][]{glad93,cham96}. 
Here, subscripts $1$ and $2$ denote the inner and outer
planets, respectively. 

\subsection{Numerical Method} \label{numer}

The previous section identified eight {\em Kepler} systems with
$\beta/\beta_{\rm crit} > $ 1.5 (Table \ref{betas}), which suggests
that these systems are most likely to have gaps between adjacent
planets that may contain additional planet(s). For these eight
systems, we numerically explore their regions of Lagrange stability to
determine zones in orbital element space that can harbor additional,
undetected planets that are stable. We use a hybrid
symplectic/Bulirsch-Stoer algorithm from an N-body integration
package, \verb Mercury  \citep{cham99}, with a timestep that sampled
1/20$^{\rm th}$ of the innermost planet's orbit. 

For each of the eight identified {\em Kepler} systems, our simulations
include the star and its two detected planets as well as 
4000$-$8000 massless\footnote[1]{ It would be ideal to perform
detailed integrations of numerous test bodies with nonzero masses,
distributed with varying distances and velocities from the star to
sample all possible orbits. However, this process would be very
computationally costly.  Since we are investigating the stability of
terrestrial-mass planets or smaller bodies, we approximate such
objects as massless test particles in our simulations. These test
particle approximations have been similarly adopted in previous
studies \citep[i.e.,][and references therein]{rive07}.} test particles
placed in between the locations of the inner and outer planets.  We
do not assign a common number of test particles to each system for
computational cost reasons.  The overall cost of the integration is a
function of the timestep and of the number of test particles.  Systems
that have inner planets with shorter orbital periods (and therefore
shorter timesteps) are assigned fewer test particles so that the
integrations may finish within a reasonable amount of time. In total, 
we integrate $\sim$8000 test particles per system except for KOI 72 
($\sim$4000 test particles) and KOI 904 ($\sim$6000 test particles).
Each system is integrated for 10$^7$ years, and test particles that
survive the length of the integration are considered stable test
particles.

For each of the eight identified {\em Kepler} systems, initial
conditions for the star and its two detected planets are given in
Table \ref{betas}. These initial conditions include the star's mass
and the known planets' masses, radii, semi-major axes, and orbital
periods. The other orbital elements of the planets--eccentricity,
inclination, argument of pericenter, longitude of the ascending node,
and mean anomaly--are currently unknown; we assume circular and
coplanar orbits and assign a random mean anomaly for the planets. The
coplanar assumption is supported by the fact that these are all
transiting planets; the larger the mutual inclination between the
planets' orbital planes, the smaller the probability that they all
transit the star \citep{rago10}. Circular and coplanar orbits have the
least angular momentum deficit and therefore are most likely to be
stable configurations \citep{lask97}.

Initial conditions for test particles are as follows.  Inclinations
$i$ are drawn from a uniform distribution ($0^{\circ} < i <
5^{\circ}$).  Previous work by \citet{liss11b} initially suggested
that {\em Kepler} multi-planet systems have low relative inclinations
with a mean of $\lesssim$5$^{\circ}$, but that number was revised to
$\lesssim$10$^{\circ}$ as our paper was undergoing revisions.
Semi-major axis $a$ and eccentricity $e$ are initially drawn from a
uniform distribution ($a_1 < a < a_2;$ $0 < e < 1$) for the first 1000
particles for each system.  Subsequent integrations of additional test
particles were randomly inserted into semi-major axis and eccentricity
bins that had few or no particles, by filling up bins with lower
eccentricity first. This ensured better coverage of semi-major axis
and eccentricity space.  All other orbital elements (argument of
pericenter, longitude of the ascending node, and mean anomaly) are
drawn randomly from a uniform distribution between 0$^{\circ}$ and
360$^{\circ}$.

This procedure is performed for each of the eight {\em Kepler} systems
identified with $\beta/\beta_{\rm crit} > $ 1.5 (shown in Table
\ref{betas}). For each system, we record each test particle's starting
orbital elements and whether it became unstable or remained stable
during the duration of the integration. Instability can be due to
ejection or collision of the test particle with another body.

% %%%%%%%%%%%%%%%%%%%%%%%%%%%%%%%%%%%
\begin{figure*}[htb]
	\centering
	\mbox{\subfigure{\includegraphics[height=2.55in]{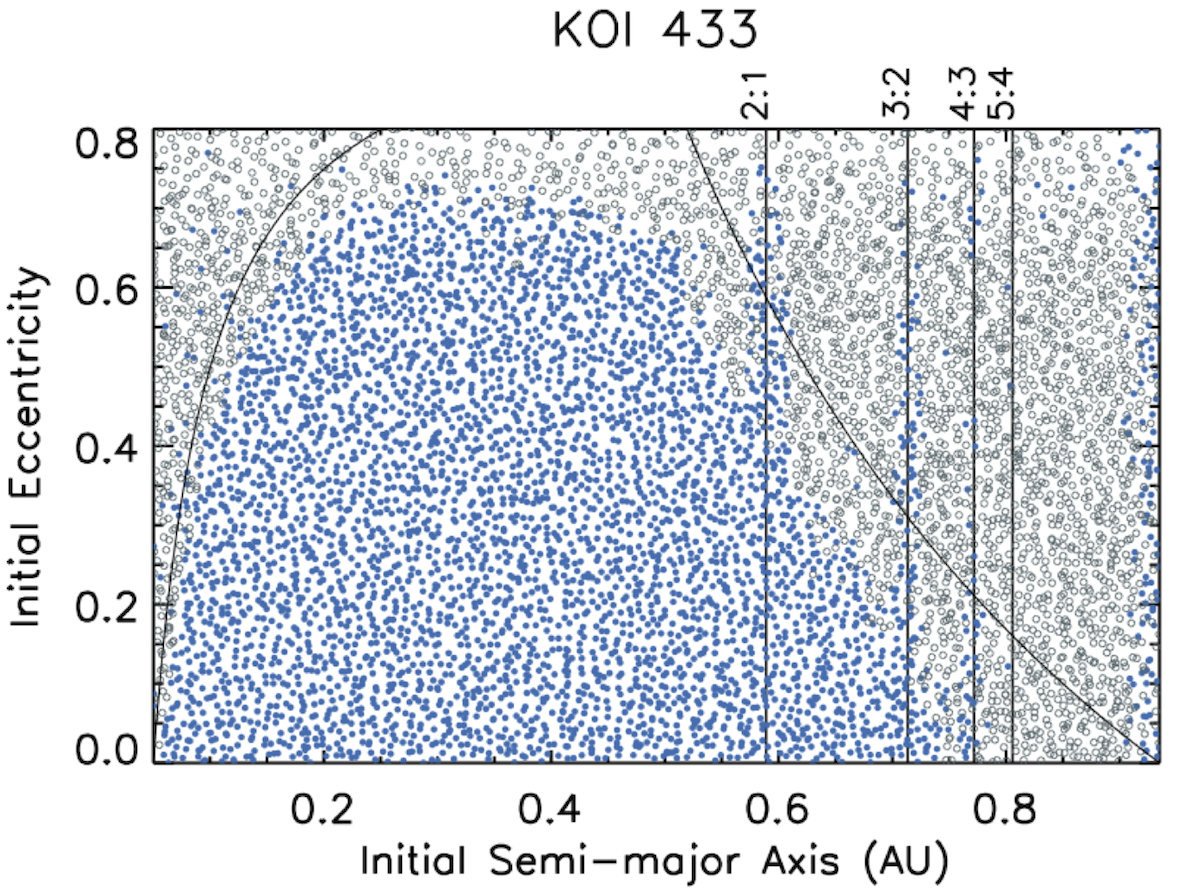}}\quad
	\subfigure{\includegraphics[height=2.58in]{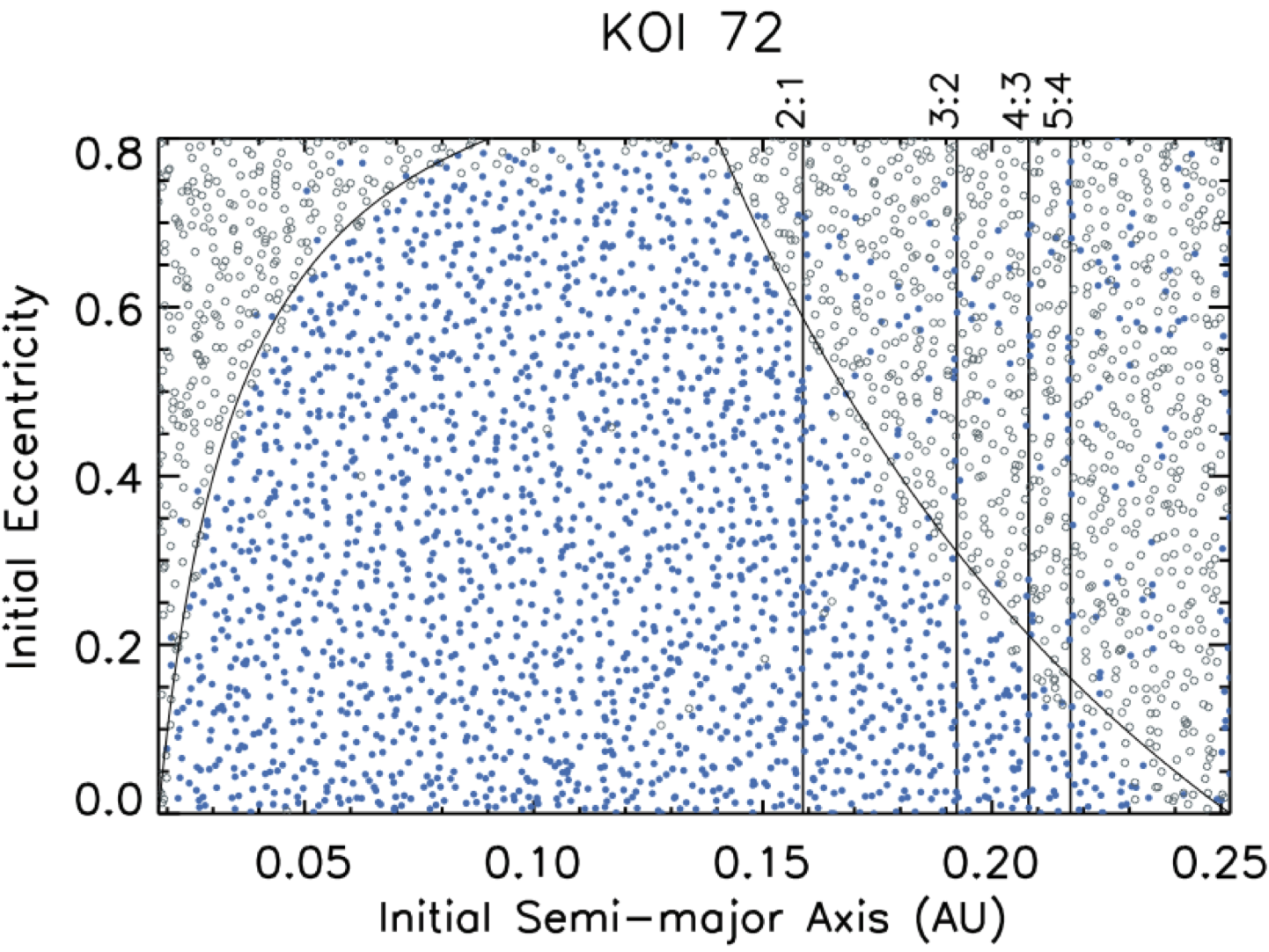}}}
	\mbox{\subfigure{\includegraphics[height=2.55in]{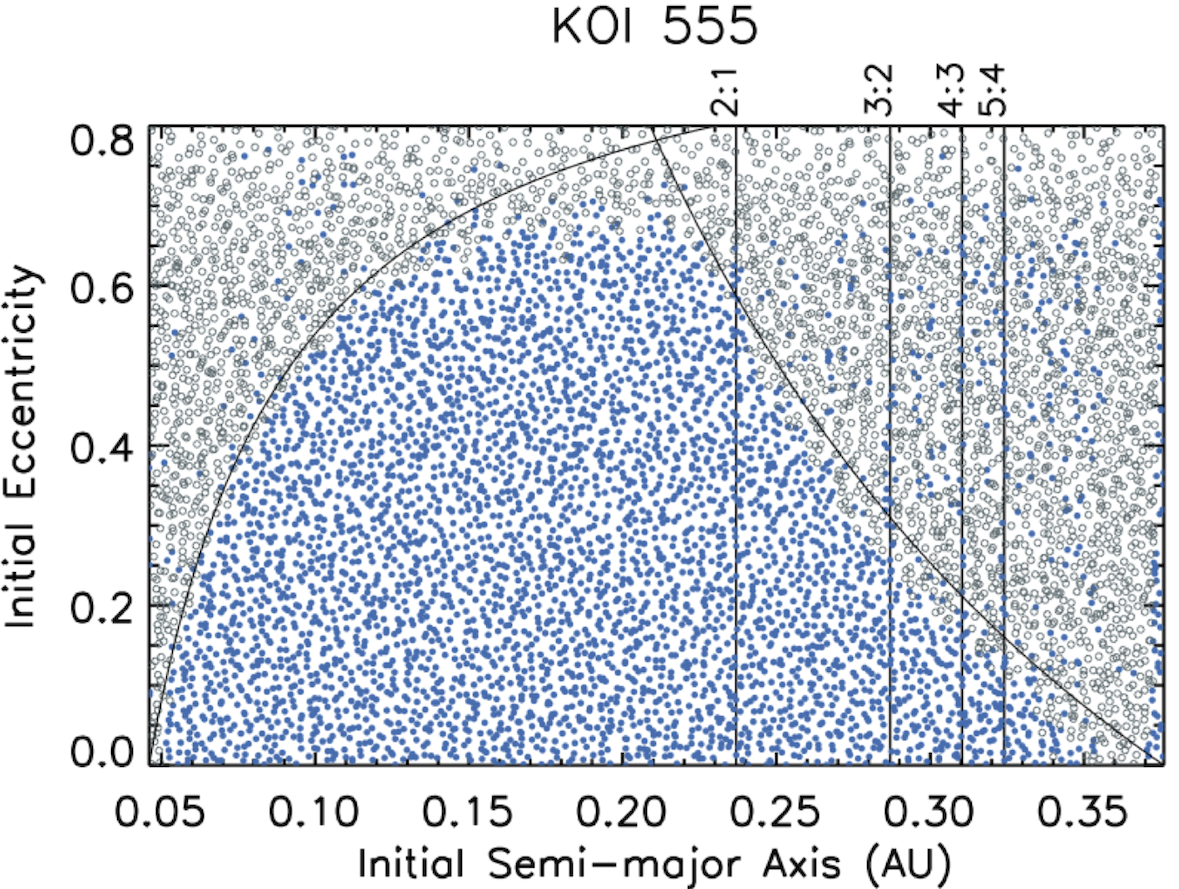}}\quad
	\subfigure{\includegraphics[height=2.58in]{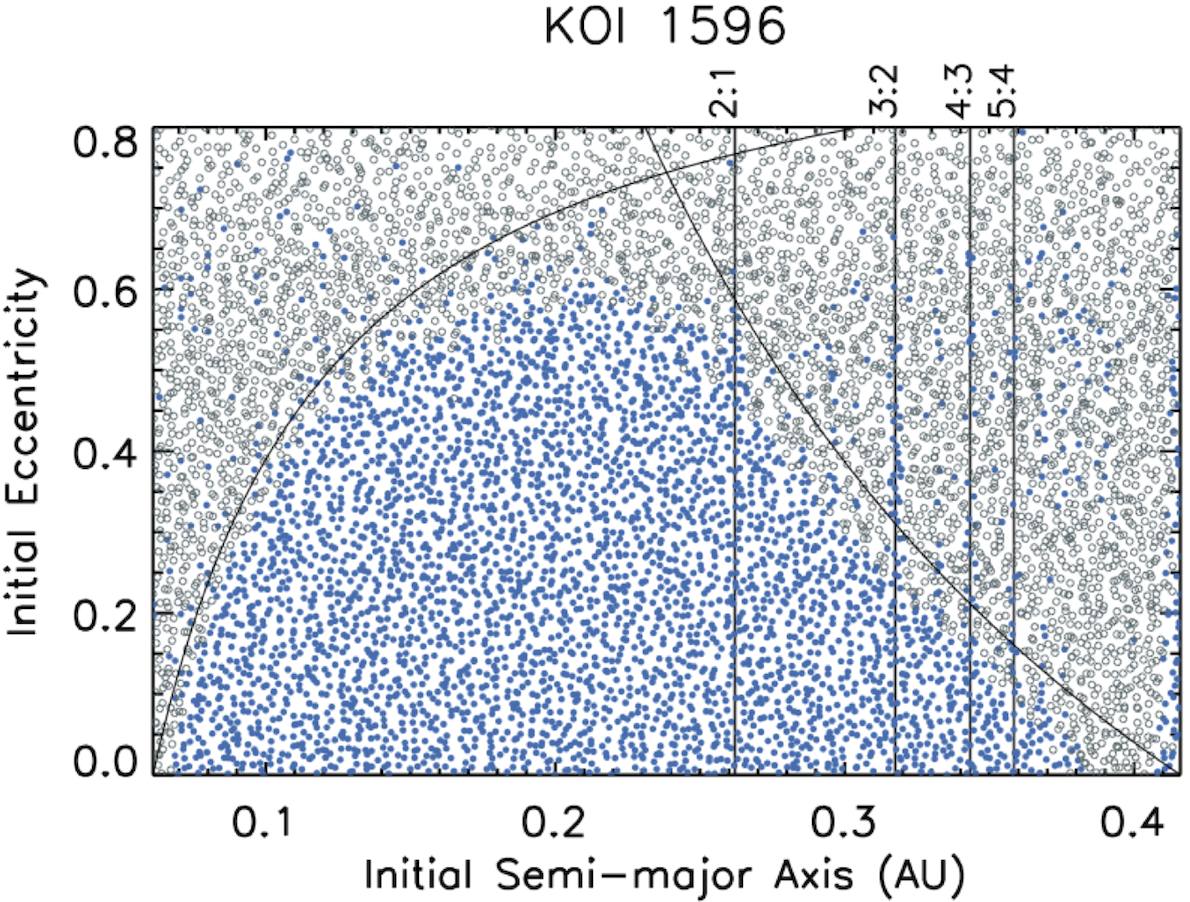}}}
	\caption{Stability maps for KOI 433, KOI 72/Kepler-10, KOI 555, and KOI 1596.
          Test particles, shown as circles, are displayed at their
          starting values of semi-major axis and eccentricity.  
          Filled blue circles are test particles that survived the
          integration length of 10$^7$ years, and unfilled gray
          circles are test particles that did not remain stable in
          that time. Black curves show the boundaries dividing planet 
	  crossing and non-planet crossing orbits. Vertical black lines 
	  represent the locations of first-order mean-motion resonances 
	  with the outer planet. The inner and outer planets are 
	  located at the left and right edges of the plot, respectively.
\label{maps1}} 
\end{figure*}
% %%%%%%%%%%%%%%%%%%%%%%%%%%%%%%%%%%%

% %%%%%%%%%%%%%%%%%%%%%%%%%%%%%%%%%%%
\begin{figure*}[htb]
	\centering
	\mbox{\subfigure{\includegraphics[height=2.55in]{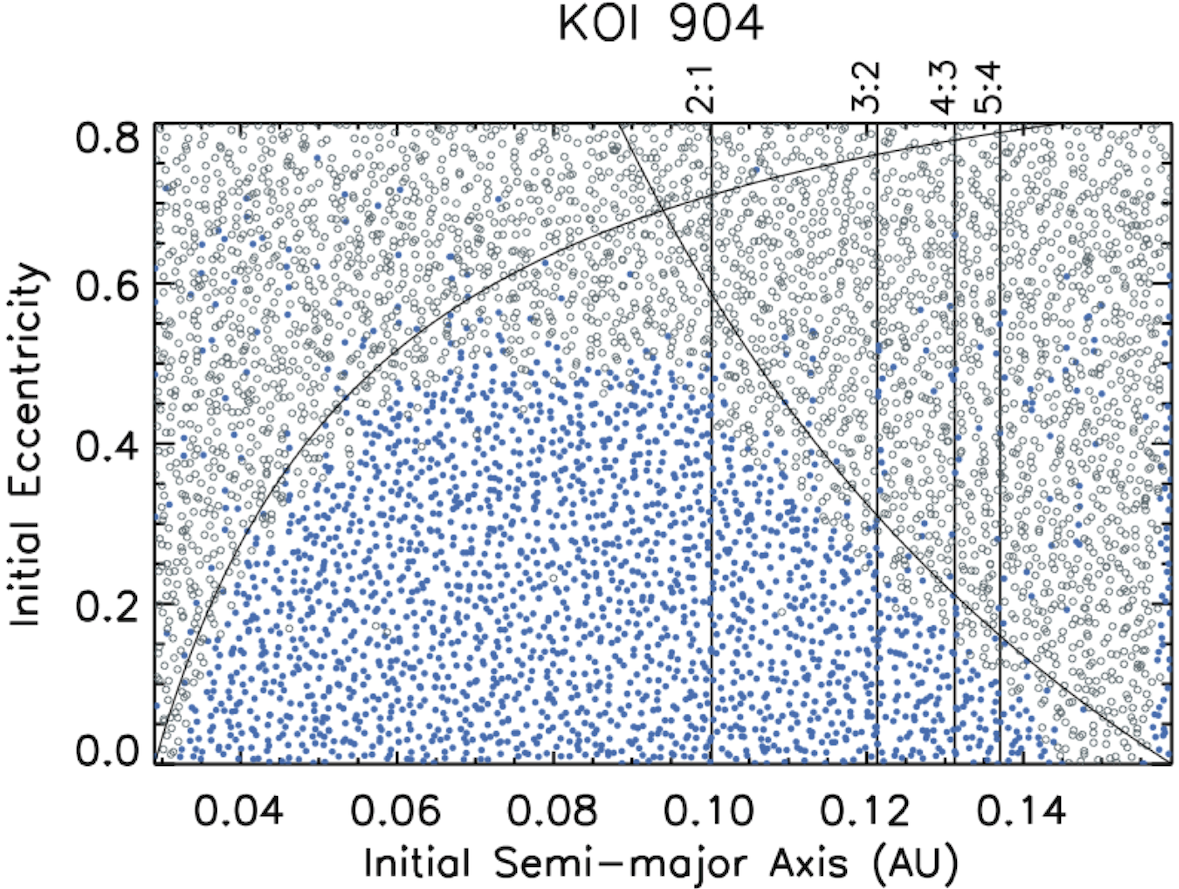}}\quad
	\subfigure{\includegraphics[height=2.58in]{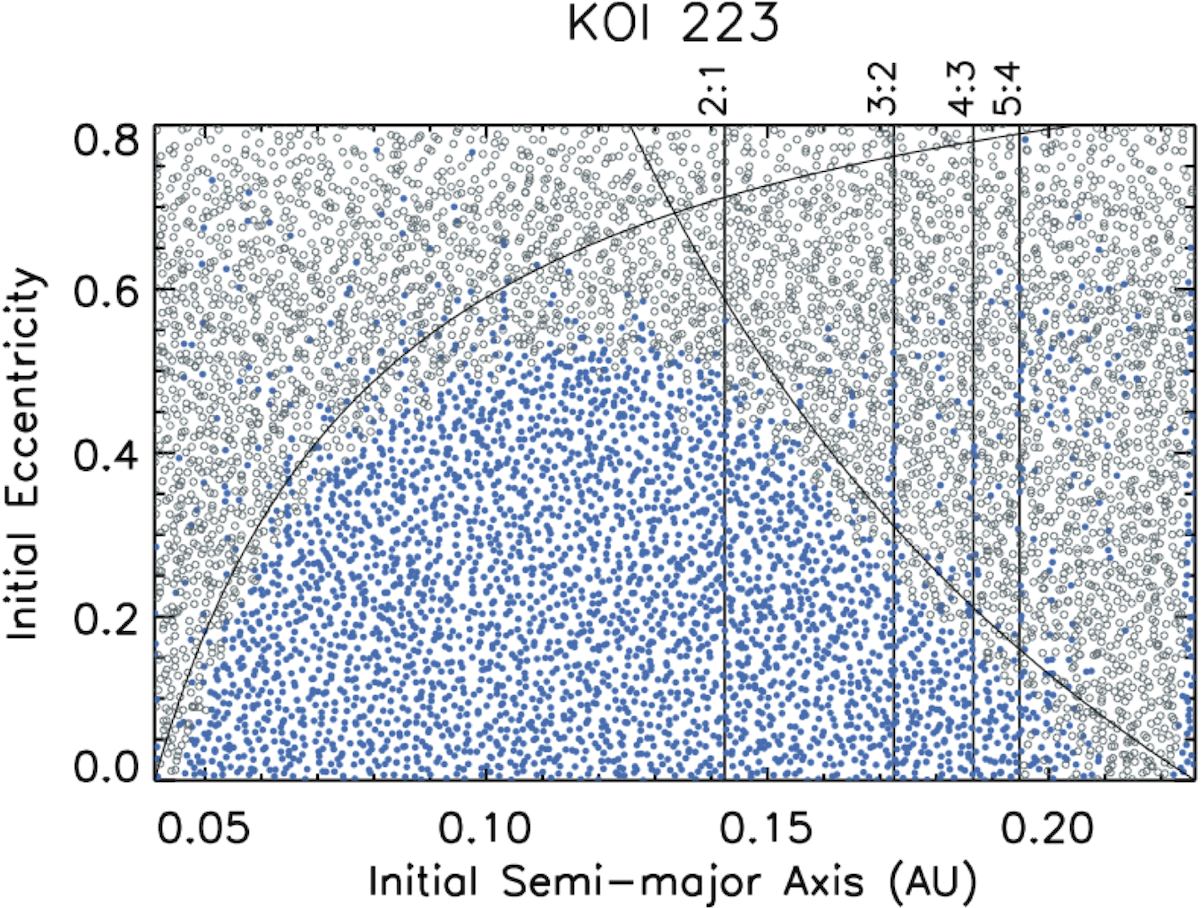}}}
	\mbox{\subfigure{\includegraphics[height=2.55in]{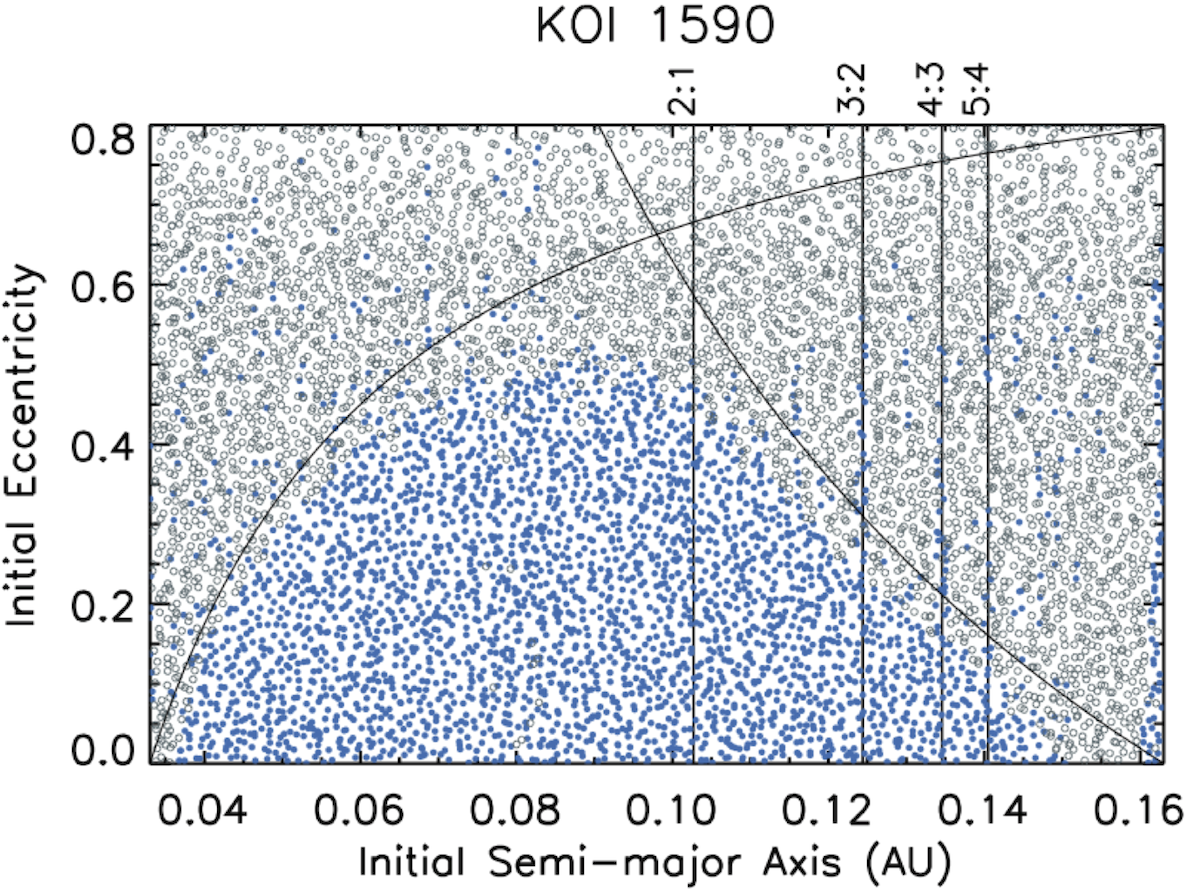}}\quad
	\subfigure{\includegraphics[height=2.58in]{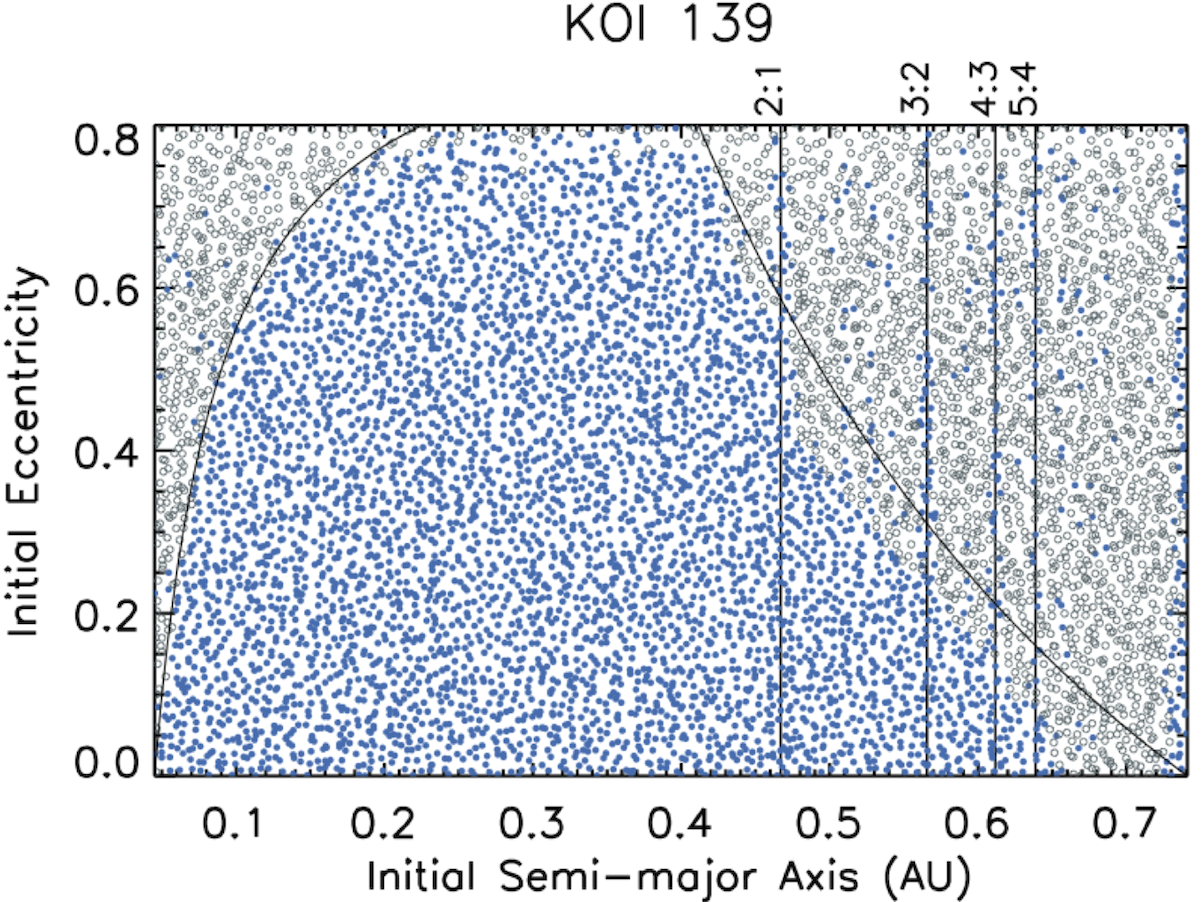}}}
	\caption{Same as Figure \ref{maps1}, except we show stability plots for 
	KOI 904, KOI 223, KOI 1590, and KOI 139. 
\label{maps3}} 
\end{figure*}
% %%%%%%%%%%%%%%%%%%%%%%%%%%%%%%%%%%%

% %%%%%%%%%%%%%%%%%%%%%%%%%%%%%%%%%%%
\section{Results} \label{results}

In this section, we describe the results stemming from long-term
N-body integrations of eight {\em Kepler} systems (Table \ref{betas}). 
All of these multi-planet systems have two known planets, and were identified in
Section \ref{analy} as potentially capable of containing an additional
planet in the regions between the inner and outer planets. For each of
these systems, we quantify their zones of stability and instability in
semi-major axis and eccentricity space by illustrating them in 
stability maps plotted in Figures \ref{maps1}$-$\ref{maps3}.

Our stability maps and results indicate that each of these planetary
systems are capable of harboring a stable low-mass body for up to
10$^7$ years in the intermediate zone between the known inner and
outer planets.  We discover broad stable regions in each planetary
system, which appear as mountain-shaped regions in Figures
\ref{maps1}$-$\ref{maps3}.  We also find additional regions of
stability outside the mountain regions, where test particles can
have stable orbits due to mean-motion resonances with the inner and
outer planets.
Strong first-order resonances with the outer planet are marked in
Figures \ref{maps1}$-$\ref{maps3}.  As for instabilities, the majority
(typically $\sim$90\%) of unstable test particles were unstable within
the first 10$^6$ years. 

Stable test particles do not show much movement in semi-major axis and
eccentricity over the course of an integration. As a result, the plots
shown in Figures \ref{maps1}$-$\ref{maps3} only show the starting
locations of test particles.  We quantify the motion of test
particles in orbital element space by computing the median of the
absolute values of the differences between initial and final values of
semi-major axis and eccentricity.  The median semi-major axis
differences range from $\sim3.8 \times 10^{-6}$ AU to $\sim2.3 \times 10^{-4}$
AU, and the median eccentricity differences range from $\sim2.4
\times 10^{-4}$ to $\sim8.5 \times 10^{-4}$.  The largest differences in
semi-major axis and eccentricity for stable particles are commonly due
to particles placed near the edge of the stability region that became
scattered off to another part of the stability region, or particles
originally not in the stability region that became scattered to an
orbit with a final semi-major axis greater than the outer planet's
semi-major axis and typically accompanied by an increase in
eccentricity.

Mean motion resonances can act as additional reservoirs of stability
outside of the mountain-shaped region. Strong first-order resonances
are plotted in Figures \ref{maps1}$-$\ref{maps3} to provide examples
of stable test particles in resonances outside of the stability
region. Many more first-order and higher-order resonances exist,
forming a thicket of resonance locations that are not drawn to reduce
confusion.  We find that the majority of stable test particles outside
the mountain are located in resonant or near-resonant periods with
that of the inner or outer planet. Test particles placed in
planet-crossing orbits (above the black curves drawn in Figures
\ref{maps1}$-$\ref{maps3}) can be stable if placed in such resonances,
which protect the test particles from close encounters with the
planets.  The role of mean motion resonances in the stability of test
particles and planets has also been previously explored \citep[e.g.,
  see][]{rive07,barn07}.

Our results show that massless test particles can stably orbit in
these stability regions for up to 10$^7$ years, and we suggest that
these stability results can be extended from massless particles to
Earth-mass planets. Spot checks performed for KOI 1596, which has a
moderate (for this sample) $\beta/\beta_{\rm crit}$ of $\sim$1.817,
indicate that an Earth-mass planet with a semi-major axis in the
middle of the stability region and with an eccentricity of zero is
stable for at least 10$^7$ years. We have also tested scenarios where
we increased the mass of the inserted planet up to a few Earth masses,
as well as cases where we inserted two, three, and four evenly-spaced,
Earth-mass planets with zero eccentricities in the main stability
region. These integrations all proved to be stable for up to 10$^7$
years in our tests for KOI 1596. As a result, it is likely that the
stability zones identified using massless test particles are
applicable to Earth-mass bodies, and that these stability zones can
potentially contain more than one Earth-mass planet.

We briefly compare our numerical results with analytical expectations.
Our stability results based on this sample of {\em Kepler} systems
indicate that two-planet systems meeting the analytical threshold
$\beta/\beta_{\rm crit} > $ 1.5 are consistent with the idea that they
can hold additional planet(s) in intermediate separations from their
host star. All eight systems investigated here had planet pairs with
$\beta/\beta_{\rm crit} > $ 1.5 and were numerically found to be
``unpacked,'' which supports previous work suggesting that additional
planets are expected to be stable in systems with $\beta/\beta_{\rm
  crit} > $ 1.5$-$2 \citep{raym09}. Since all eight {\em Kepler}
systems we investigated with $\beta/\beta_{\rm crit} > $ 1.5 are
unpacked, we expect that there may also be additional systems with
$\beta/\beta_{\rm crit}$ less than 1.5 that are also unpacked (see
next section).

% %%%%%%%%%%%%%%%%%%%%%%%%%%%%%%%%%%%
\section{Planetary Spacing Determines Extent of Stability Region} \label{mountain}

We now describe in greater detail the shapes and sizes of the
mountain-shaped stability regions observed in Figures 
\ref{maps1}$-$\ref{maps3}.
In particular, we discuss relationships between the spacing between
two planets and the extent of the stability region in-between the
planets.

The stable regions in each planetary system include a mountain-shaped
stability peninsula as well as narrow strips of stability due to
mean-motion resonances (e.g.,~see Figures \ref{maps1}$-$\ref{maps3}).
The large mountain-shaped stability region has a shape common to all
of the planetary systems, because it is sculpted on the left and right
flanks by specific semi-major axis and eccentricity values that
delineate planet-crossing orbits.  Mathematically, the mountain's left
slope is shaped by $a_1 = a(1-e)$, where $a_1$ is the inner planet's
semi-major axis. The mountain's right slope is shaped by $a_2 =
a(1+e)$, where $a_2$ represents the outer planet's semi-major
axis. These orbit-crossing boundaries are shown as black curves in
each stability plot in Figures \ref{maps1}$-$\ref{maps3}.  The actual
stability boundaries (left and right flanks of the mountain-shaped
stability region) do not extend all the way to the black curves.  This
is explained by close approach effects: test particles that are not
initially on planet-crossing orbits can become unstable if they make
sufficiently close approaches to the existing planets.  The critical
distance from the planet-crossing boundary at which this can occur is
similar to the half-width of a planet's ``feeding zone'' in which
planetesimals may impact the planet, which can be estimated at about
$\sim$2.3 Hill radii for circular orbits \citep[i.e.,][]{gree91}. The
results of our simulations show similar distances between the
planet-crossing curve and the actual slope of the mountain.

The maximum height of each mountain-shaped stability region is
constrained by the semi-major axes of the inner and outer planets. The
maximum eccentricity allowed is determined by the intersection of the
orbit-crossing boundaries, $a_1 = a(1-e)$ and $a_2 = a(1+e)$, and
serves as an upper limit to the maximum possible height of the
stability mountain. Since we assume that the two known planets in each
system have orbital eccentricities of zero, the intersection of the
curves occurs at a semi-major axis of $(a_1+a_2)/2$ and an
eccentricity of
\begin{align} \label{emax}
	e_{\rm max} = 1 - \dfrac{2a_1}{a_1+a_2},
\end{align}
which is the maximum possible eccentricity $e_{\rm max}$ of the
stability mountain. As evident in Figures \ref{maps1}$-$\ref{maps3},
the actual peak of the stability region is not the same as the $e_{\rm max}$.  

\begin{figure}[htb]
	\centering
	\includegraphics[height=2.4in]{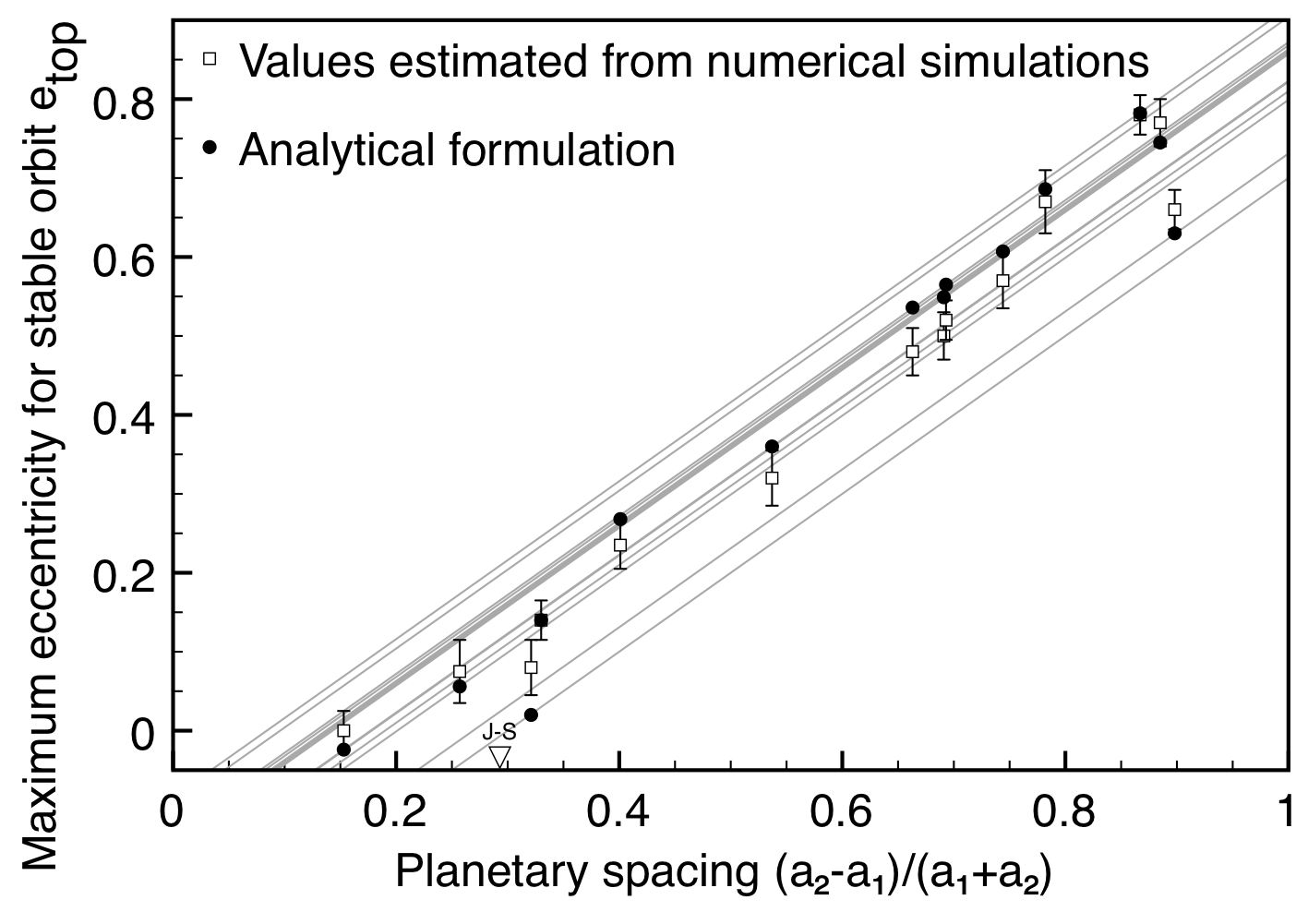}
	\caption{Maximum eccentricity for a stable test particle orbit 
	at a semi-major axis $(a_1+a_2)/2$ between two existing planets. 
	The unfilled squares represent 
        estimates of $e_{\rm top}$ with their uncertainties, and the filled
        circles and gray lines represent values of $e_{\rm top}$
        computed from Equation (\ref{etopeqn5}) for various planetary
        systems (eight systems listed in Table \ref{betas} plus six
        additional systems for a larger sample). For comparison, the
        inverted triangle shows the planetary spacing between Jupiter
        and Saturn (note that we only considered two-planet systems
        with circular orbits, and our results may not be applicable to
        systems with greater multiplicity of planets or non-circular
        orbits).
\label{etop}} 
\end{figure}

The actual height or peak of each mountain-shaped stability region can
be computed as follows.  Consider a test particle located between the
inner and outer planets.  In order to remain stable, this particle
cannot enter a zone of dynamical influence surrounding each planet.
We measure this exclusion zone as a certain number $c_i$ of Hill radii
$R_{Hi}$, where $R_{Hi} = (M_i/(3M_*))^{1/3}a_i$ and the subscript
$i=1,2$ refers to the inner and outer planets, respectively.
Therefore, a stable test particle's pericenter $q = a(1-e)$ and
apocenter $Q = a(1+e)$ distances must obey
\begin{align} \label{etopeqn1}
	q = a(1-e) > a_1 + c_1R_{H1}
\end{align}
\begin{align} \label{etopeqn2}
	Q = a(1+e) < a_2 - c_2R_{H2},
\end{align}
where the inner and outer planets are assumed to have circular orbits.
We label the maximum stable eccentricity as $e_{\rm top}$ (flat top of
the mountain) and consider the midpoint between the two planets
$(a_1+a_2)/2$.
We can rewrite Equations (\ref{etopeqn1})$-$(\ref{etopeqn2}) for
particles on the edge of stability/instability as
\begin{align} \label{etopeqn3}
	\dfrac{(a_1+a_2)}{2} (1-e_{\rm top}) = a_1 + c_1R_{H1}
\end{align}
\begin{align} \label{etopeqn4}
	\dfrac{(a_1+a_2)}{2} (1+e_{\rm top}) = a_2 - c_2R_{H2}.
\end{align}
If we subtract the two equations from each other and solve for $e_{\rm top}$, we obtain
\begin{align} \label{etopeqn5}
	\nonumber e_{\rm top} &= \dfrac{a_2 - a_1 - c_1R_{H1} - c_2R_{H2}}{a_1 + a_2} \\
	&= - \dfrac{c_1R_{H1} + c_2R_{H2}}{a_1+a_2} + \dfrac{a_2 - a_1}{a_1 + a_2}.
\end{align}
We empirically determine $c_1$ and $c_2$ by fitting Equation
(\ref{etopeqn5}) in a least-squares sense to values of $e_{\rm top}$
measured from Figures \ref{maps1}$-$\ref{maps3}.  We find $c_1$ =
19.733 and $c_2$ = 4.1877.  Comparison between computed values of
$e_{\rm top}$ (using Equation (\ref{etopeqn5})) and the measured
values of $e_{\rm top}$ (from Figures \ref{maps1}$-$\ref{maps3}) is
shown in Figure \ref{etop}, which illustrates the maximum stable
eccentricity $e_{\rm top}$ as a function of planetary spacing for a
range of planetary masses.

% %%%%%%%%%%%%%%%%%%%%%%%%%%%%%%%%%%%

% %%%%%%%%%%%%%%%%%%%%%%%%%%%%%%%%%%%

% %%%%%%%%%%%%%%%%%%%%%%%%%%%%%%%%%%%
\begin{figure}[htb]
	\centering
	\includegraphics[height=2.4in]{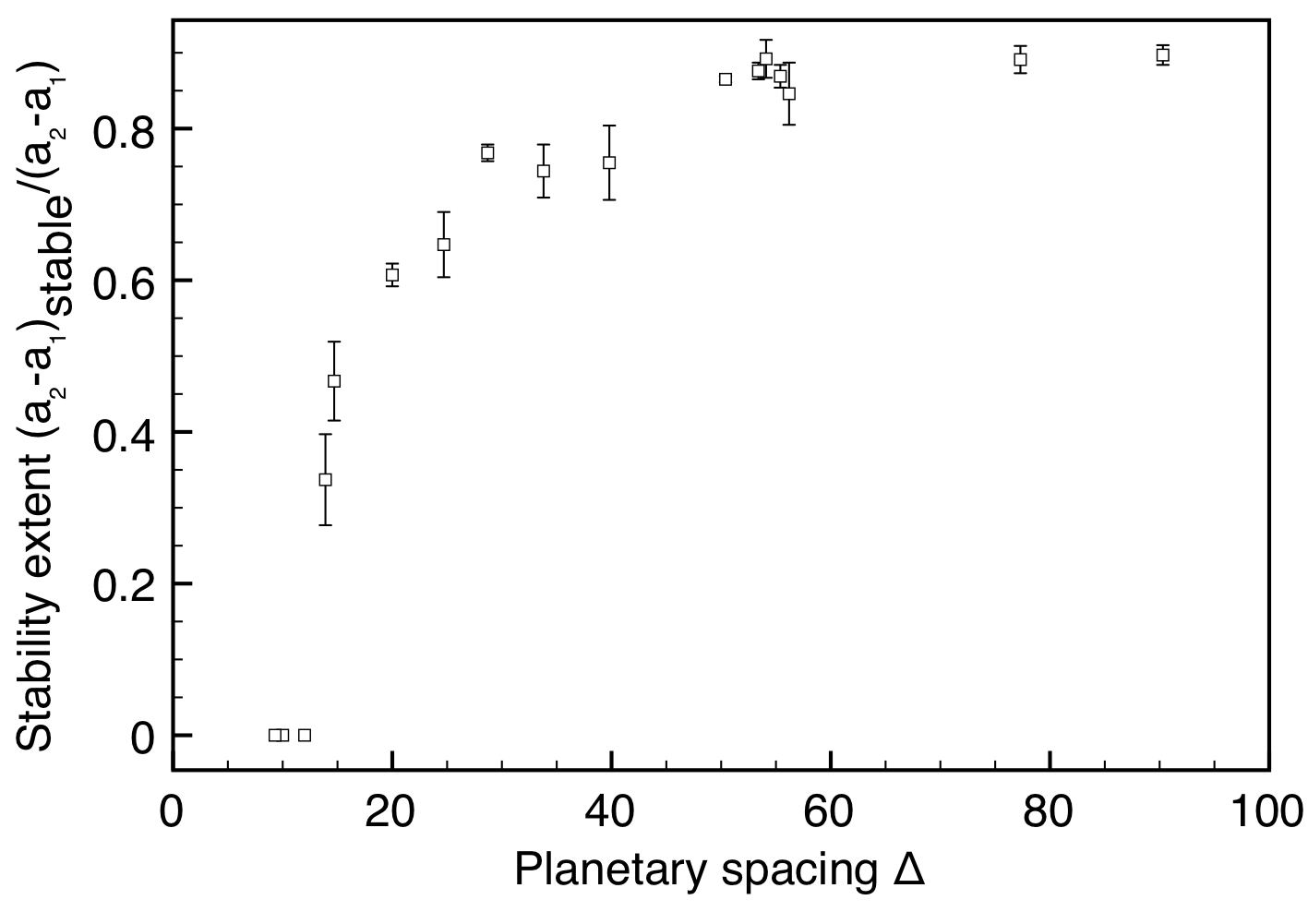}
	\caption{Fraction of planetary separation $(a_2-a_1)$ with stable 
	test particles at $e = 0$, as a function of planetary spacing criterion $\Delta$.
	The unfilled squares represent the results from numerical 
	simulations in this study (eight planetary systems from Table 
	\ref{betas} plus nine additional systems for a larger sample) with 
	error bars representing measurement uncertainties.
\label{delta}} 
\end{figure}
% %%%%%%%%%%%%%%%%%%%%%%%%%%%%%%%%%%%

The width of a stability mountain's base (where $e = 0$) can be
related to the dynamical spacing criterion $\Delta$ between two
planets (see Equations (\ref{deltaeqn1}) and (\ref{deltaeqn2})).  A system
with two planets in a circular and coplanar state satisfies Hill
stability (orbits do not cross) if $\Delta$ is greater than $2\sqrt3$,
or $\sim$3.46 \citep{glad93}. The stability of systems with more than
two planets are less well-characterized and are commonly determined
using numerical calculations.  Estimates of the width $(a_2-a_1)_{\rm
  stable}$ of each stability mountain's base at $e = 0$ (ignoring the
effects of resonances as much as possible) are related to $\Delta$
(Figure~\ref{delta}).  We do not find any 
stability regions for planetary systems with $\Delta \lesssim 10$.

We generalize the results shown in Figure \ref{delta} to a broader
context. From this figure, we can determine a critical value of
$\Delta$ that divides two-planet systems with stable versus no stable
regions. This cross-over occurs in the range $\Delta_{\rm crit} =
10-15$.  Accordingly, we suggest that two-planet systems similar to
those explored in this paper cannot have extensive stability zones if
their separations have $\Delta$ less than 10.  Similarly, we predict
that stable regions can exist in systems with $\Delta$ greater than
15. In the February 2011 {\em Kepler} release \citep{boru11,liss11b},
95 out of a total of 115 two-planet systems have $\Delta > 15$, or
82.6\% of all two-planet systems in this sample can potentially harbor
stability zones within the known planets. The results discussed here
and illustrated in Figure \ref{delta} are consistent with previous
studies.  \citet{cham96} numerically studied coplanar and circular
configurations of 3, 5, 10, and 20 planet systems, and found no stable
systems with planetary spacing of $\Delta < 10$. More recently,
\citet{smit09} examined the packing density of systems with 3, 5, and
9 Earth-mass planets in circular and coplanar orbits with planets
equally spaced in terms of $\Delta$. They conducted long-term
numerical integrations up to 10 billion years, and demonstrated that
3-planet systems are stable when the spacing between neighboring
planets is roughly $\Delta \sim 7$. Other previous results on spacing
between planets or protoplanets include the typical $\sim$10 Hill
radii spacing between neighboring protoplanets, as seen in simulations
of protoplanetary accretion from a swarm of planetesimals
\citep{koku98,koku00,koku02}.

\section{Scope and Limitations of Our Results} \label{scope}

Given the large amount of possible parameter space that can be
explored in stability studies, we summarize the limitations and scope
of results stemming from this paper. We also discuss any other
assumptions and considerations that may change our results.

{\em Sample.} We solely investigated multi-planet systems announced by
the {\em Kepler} team in the February 2011 release of candidate
systems \citep{boru11}.  No other planetary systems were
considered. Therefore, our sample has the same biases as any {\em
  Kepler} detection, including the observational preference towards
short-period planets given the transit detection method.  Our study is
also limited to {\em Kepler} systems for which there are two known
planets, and we do not investigate the dynamical spacing in systems
with greater multiplicity of planets.

{\em Masses.} The planetary masses are typically not known for these
KOI systems. We estimated masses using {\em Kepler}-measured planetary
sizes with a power law (Table \ref{betas}, \citet{liss11b}) obtained
from fitting to Earth and Saturn. However, the densities and true
masses of {\em Kepler} planets can be different from these
assumptions, which could change our results.

{\em Eccentricities.} Eccentricity is another important dynamical
parameter that is not known for most {\em Kepler} multi-planet
systems. We have assumed zero eccentricities for the known planets in
our numerical calculations, and this assumption is consistent with the
expected tidal circularization of close-in planets. For the only
confirmed planetary system in our sample, KOI 72 or Kepler-10,
photometry and radial velocity data 
suggest that Kepler-10b has zero eccentricity \citep{bata11}.
Non-zero eccentricities of the planets in our sample, if present,
would change the locations of stability regions of test particles.

{\em Inclinations.} We assumed zero inclinations between the orbits of
known planets in our sample as well as low inclinations up to
$\sim$5$^{\circ}$ for test particles. Consequently, our results can
only be applied to systems that are relatively coplanar.  The
assumption of coplanarity or near-coplanarity is reasonable for
multi-planet systems discovered by the transit technique at the heart
of the {\em Kepler} mission, given that the inclination dispersion of
these systems appears to have mean of $\lesssim 10^{\circ}$.

{\em Integration time.} We integrated test particles for a time span
of 10$^7$ years due to CPU time limitations, but more accurate
modeling can be obtained by using a longer integration time
period. There may be test particles that are stable over 10$^7$ years
but not over longer timescales, although our simulations show that
$\sim$90\% of particles unstable in 10$^7$ years were unstable within
the first 10$^6$ years.

% %%%%%%%%%%%%%%%%%%%%%%%%%%%%%%%%%%%
\section{Conclusion} \label{conclusion}

The ``packed planetary systems'' model advocates the idea that all
planetary systems are formed to capacity. To test this hypothesis, we
investigated the packing density of {\em Kepler} candidate two-planet
systems from the first four and a half months of the mission. Through
numerical calculations, we determined whether regions of stability
exist between known planets with wide separations, i.e.,~in systems
that seemed the most unpacked based on how well they satisfy Hill
stability.
Discovery of a stable region suggests that a low-mass body could be
present in the gap, which would then bring the system to a more
``packed'' state. With time, such predictions will be shown to be
correct or incorrect, allowing us to gauge the success of this model.

We performed detailed numerical simulations of eight, two-planet {\em
 Kepler} systems, selected using an analytical $\beta/\beta_{\rm
 crit}$ stability criterion. In addition to the known planets, we
included 4000$-$8000 test particles per planetary system, allowing
both circular and non-circular, and coplanar and non-coplanar
orbits. These test particles are good proxies for low-mass bodies such
as terrestrial planets as well as small bodies such as asteroids or
dwarf planets. We integrated all bodies for 10$^7$ years; we defined
stable particles as those that remained stable during the length of
the integration and unstable particles as particles that experienced a
collision or ejection.

Our results (Figures \ref{maps1} to \ref{maps3}) indicated that all of
the planetary systems investigated here (KOIs 433, 72, 555, 1596, 904,
223, 1590, and 139) can pack additional, yet-undetected bodies in the
identified stable locations. We also discussed relationships relating
dynamical spacing between known planets and the extent of the
inter-planet stability region. We derived an analytical relationship 
relating the largest possible eccentricity of a stable test particle 
to the semi-major axes and Hill radii of the two planets surrounding 
the particle. We also demonstrated that $\Delta$, the
separation between two planets in units of their mutual Hill radii,
can be a reasonable predictor of whether or not stability regions can
exist between planets.  The cut-off occurs at a critical $\Delta$
between 10 and 15.  We suggest that planets with separation $\Delta <
10$ are unlikely to host 
extensive stability regions.  Based on this $\Delta=10-15$ cut-off, we
suggest that about 95 out of a total of 115 two-planet systems in the
February 2011 {\em Kepler} sample may have sizeable stability regions.

\acknowledgments
We thank the referee for useful comments.

% %%%%%%%%%%%%%%%%%%%%%%%%%%%%%%%%%%%
\bibliographystyle{apj}
\bibliography{packing}

\end{document}